\documentclass{iopart}

\usepackage{iopams}
\usepackage{graphicx}

\begin{document}

\title[Density functional for hard sphere mixtures]
{Close to the edge of Fundamental Measure Theory:
density functional for hard sphere mixtures}

\author{Jos\'e A Cuesta\dag, Yuri Mart\'{\i}nez-Rat\'on\dag\ and \\
Pedro Tarazona\ddag}

\address{\dag\ Grupo Interdisciplinar de Sistemas Complicados (GISC),
Departamento de Mate\-m\'a\-ti\-cas, Universidad Carlos III de Madrid, Avda.\ 
de la Universidad, 30, E-28911 Legan\'es, Madrid, Spain \\
\ddag\ Departamento de F\'{\i}sica Te\'orica de la Materia Condensada,
Universidad Aut\'onoma de Madrid, E-28049 Madrid, Spain.
}

\begin{abstract}
We analyze the structure of the Fundamental Measure Theory
for the free energy density functional of hard sphere mixtures.
A comparative study of the different versions of the theory,
and other density functional approaches, is done in terms of 
their generic form for the three-points direct correlation 
function, which shows clearly the main advantages and 
problems of the different approximations. A recently
developed version for the monocomponent case is extended
to mixtures of hard spheres with different radii, and a new
prescription is presented to obtain the exact dimensional 
crossover of those mixtures in the one-dimensional (1D) limit.
Numerical results for planar wall-fluid interfaces and for
the 1D fluid are presented.
\end{abstract}

%\submitto{\JPCM}

%\maketitle

\eqnobysec

\section{Introduction}

In the recent developments of density functional (DF) approximations for 
the free energy of hard molecules the Fundamental Measure Theory (FMT),
pioneered by Rosenfeld \cite{Yasha} and explored in different directions 
by other authors \cite{others,White,cubos,schmidt}, 
stands at a prominent position both 
because of its peculiar functional structure and because of the 
combination of success and pitfalls obtained with its
different versions. In this paper we analyze the generic mathematical 
structure of the different density functional approximations
and discuss its implication for the generalization of a recent version
of FMT to mixtures of hard spheres with different radii.  

The theories developed to approximate the interaction part of the 
free energy density functional $\Phi[\rho]\equiv \beta(F[\rho]-
F_{\rm id}[\rho])$, with $\beta=(k_BT)^{-1}$,
beyond the simplest local density and square gradient approximations,
are all based on a similar scheme: the non-local dependence on the 
density distribution $\rho({\bf r})$ appears through `averaged' or
`weighted' densities defined as convolutions
\begin{equation}
\bar{\rho}_k({\bf r})=\int \rmd{\bf r}'\,\rho({\bf r}') \ w_k({\bf r}-{\bf r}'),
\label{barra}
\end{equation}  
with several ($k=0,1,..$) weight functions, $w_k({\bf r}-{\bf r}')$, 
which depends only 
on the relative position of two points. The density functional
$\Phi[\rho]$ is then written as the volume integral of a function of
$\rho({\bf r})$ and/or the weighted densities $\bar{\rho}_k({\bf r})$. 
Such structure of the usual approximations for $\Phi[\rho]$
contrasts with the generic exact density expansion of the free energy,
\begin{eqnarray}
\fl \Phi[\rho]={1 \over 2}
\int \rmd{\bf r}_1\,\rho({\bf r}_1)\int \rmd{\bf r}_2\,\rho({\bf r}_2) 
f({\bf r}_{12}) \nonumber \\
\lo{\displaystyle +{1 \over 6}\int}\rmd{\bf r}_1\,\rho({\bf r}_1)
\int \rmd{\bf r}_2\,
\rho({\bf r}_2)\int \rmd{\bf r}_3\,\rho({\bf r}_3) 
f({\bf r}_{12})f({\bf r}_{23})f({\bf r}_{31})+\Or(\rho^4),
\label{virial} 
\end{eqnarray} 
in terms of the Mayer function $f({\bf r})=1-\rme^{-\beta\phi({\bf r})}$
for any pair potential energy $\phi({\bf r})$ (we have denoted
${\bf r}_{ij}\equiv{\bf r}_i-{\bf r}_j$).
The first term in \eref{virial} has the simplest structure of the
density functional approximations, with a basic weight function 
$w_0({\bf r})$ proportional to $f({\bf r})$,
used to convolute $\rho({\bf r})$ at two different points.
However, the second and followings terms in \eref{virial} cannot be 
reproduced within the generic forms used in the DF approximations. In 
the diagrammatic expansion used in the theory of liquids 
\cite{Hansen}, the free energy is given by the irreducible diagrams 
(with Mayer function links), which cannot be evaluated as a single volume 
integral of simpler factors, while the generic form of the DF 
approximations includes only reducible diagrams with $w_k({\bf r})$
links.

The second and third functional derivatives of \eref{virial}
give the exact series expansions for the 
pair and triplet direct correlations functions respectively,
which have the corresponding density expansions,
\begin{equation}
\fl -c^{(2)}({\bf r}_1,{\bf r}_2) = 
{{\delta^2\Phi[\rho]}\over{\delta\rho({\bf r}_1)\delta\rho({\bf r}_2)}}
=f({\bf r}_{12}) + f({\bf r}_{12}) \int \rmd{\bf r}_3\,\rho({\bf r}_3)
f({\bf r}_{23})f({\bf r}_{31}) 
+\Or(\rho^2),
\label{c2virial}
\end{equation}
and
\begin{equation}
\fl -c^{(3)}({\bf r}_1,{\bf r}_2,{\bf r}_3)={{\delta^3 \Phi[\rho]}\over
{\delta\rho({\bf r}_1)\delta\rho({\bf r}_2)\delta\rho({\bf r}_2)}}
=f({\bf r}_{12})f({\bf r}_{23})f({\bf r}_{31})+\Or(\rho),
\label{c3virial}
\end{equation}
again made of irreducible diagrams, 
with two and three open (not integrated) points respectively. 
The first term in \eref{c2virial} is reproduced by most DF approximations,
but the first term in \eref{c3virial}, and hence the second term in
\eref{c2virial} for arbitrary density distributions, cannot be 
reproduced within those DF forms.  This contradiction 
between the exact forms and the usual
choices for the mathematical structure of the DF approximations
is forced by the computational 
cost of evaluating irreducible terms. The inclusion of
kernel functions with more than two centers in the convolutions
of $\rho({\bf r})$ would make it difficult to use the DF approximation
for practical purposes.

The weighted density approximation (WDA) \cite{ADA,WDA} gives a 
good example of the compromise between accuracy and practical 
computability in DF approximations. Although the advanced versions
of this DF approximation use a density dependent weight $w(r,\bar{\rho})$, 
which may appear to be beyond the scope of \eref{barra}, for the present 
discussion it is fully equivalent to take the zeroth and first
order contributions in the density expansion
$w(r,\rho)=w_0(r)+w_1(r) \rho +\cdots$.  Then, the approximation for 
$\Phi[\rho]$ up to third order in $\rho({\bf r})$ is 
\begin{eqnarray}
\fl \Phi_{\rm WDA}[\rho]={1 \over 2}\int \rmd{\bf r}_1\,\rho({\bf r}_1)\int 
\rmd{\bf r}_2\,\rho({\bf r}_2)f({\bf r}_{12}) \nonumber \\
\lo{\displaystyle +{1 \over 6}\int \rmd{\bf r}_1}\,\rho({\bf r}_1)
\int \rmd{\bf r}_2\,\rho({\bf r}_2)\int \rmd{\bf r}_3\,\rho({\bf r}_3) 
f({\bf r}_{12})[f({\bf r}_{23})+w_1({\bf r}_{23})]+\Or(\rho^4),
\label{virialWDA}
\end{eqnarray}
where the first term reproduces directly the first term in \eref{virial},
while the second term tries to mimic the irreducible kernels as 
reducible combinations of Mayer function links, $f({\bf r})$, with a new
link function, $w_1({\bf r})$, defined to recover the direct correlation 
function of a bulk fluid, i.e.\ to recover the exact expansion 
\eref{c2virial} up to first order in the uniform density 
$\rho({\bf r})=\rho_0$. This requirement leads to a function 
$w_1({\bf r})$ which exceeds the range of the Mayer function
with an oscillatory tail structure.

Thus, the WDA for the hard sphere (HS) fluid may be implemented \cite{WDA}
to recover the direct correlation of the bulk fluid,
$c(r_{12},\rho_0)=c^{(2)}({\bf r}_1,{\bf r}_2)|_{\rho({\bf r})=\rho_0}$,
given by the Percus-Yevick approximation \cite{PY,Hansen}, with the range of 
a HS diameter, $\sigma$; however the function $c^{(2)}({\bf r}_1,
{\bf r}_2)$, evaluated for a non-uniform density distribution, would 
exceed that range because of the convolution of the Mayer function with 
the oscillatory function $w_1({\bf r})$. Similarly, the zeroth order 
term of the triplet direct correlation $c^{(3)}({\bf r}_1,{\bf r}_2,
{\bf r}_3)$ would exceed the range $r_{ij}=|{\bf r}_i-{\bf r}_j|\le
\sigma$ which the exact form \eref{c3virial} has for any combination 
of its variables. This being so, the WDA and similar approximations 
give a blurred representation of the exact non-local dependence of 
$\Phi[\rho]$ on $\rho({\bf r})$, where the sharp step-like dependence 
on the HS Mayer function in \eref{virial}--\eref{c3virial} is replaced by 
an easier to compute but smoother functional. The extension of
this type of DF to mixtures of HS with different radii $R_i$ ($i=1,\dots,m$)
is difficult, because they require $m(m+1)/2$ weight functions
to reproduce the Mayer functions $f_{ij}(r)$ for any pair of molecules,
and the `extended' weights required to mimic the irreducible diagrams
for any triplet ($i,j,k=1,\dots,m$) proliferate into an embarrassing large
number of possible combinations for the functional $\Phi[\rho]$.

The FMT also approximates $\Phi[\rho]$ for hard sphere systems
in terms of convolutions \eref{barra} of $\rho({\bf r})$, but with 
weight functions having the range of the molecular radius, $R$, rather
than the range of the Mayer function, $\sigma=2R$. The first advantage
of such a choice is a natural extension to HS mixtures, since the
number of different weight functions is going to be proportional to
the number of species, and (at least in the original version)
they only appear in a fixed number of linear combinations. The first 
one is the local packing fraction, which for a mixture of
($i=1,\dots,m$) species of hard spheres with radii $R_i$ and
density distributions $\rho_i({\bf r})$, is defined as
\begin{equation}
\eta({\bf r})=\sum_{i=1}^m\int \rmd{\bf r}'\,\rho_i({\bf r}') 
\Theta(R_i-|{\bf r}-{\bf r}'|),
\label{eta}
\end{equation}    
with the Heaviside step function $\Theta(x)$. This definition of
$\eta({\bf r})$ is the direct extension to inhomogeneous systems 
of the usual packing fraction $\eta=(4\pi/3)\sum_i R_i^3\rho_i$,
which plays a most relevant role in the best approximations for the 
equation of state of bulk fluid HS mixtures. Also, $\eta({\bf r})$
is a key ingredient in the exact density functional for hard rods
\cite{Percus} in one dimension (1D), and it has the appealing 
interpretation 
of being the probability that point ${\bf r}$ happens to be inside 
a sphere. Despite these facts, the local packing fraction cannot
appear in any DF based on the Mayer function and its convolutions,
like the WDA in \eref{virialWDA}, because all the weight functions have
at least the range of two added molecular radii. The problem to include
$\eta({\bf r})$ in the functional dependence of $\Phi[\rho]$ is to
recover the Mayer function in the lowest order terms in \eref{virial}
and \eref{c2virial}, from kernels with  half the range. In the exact
density functional for hard rods in 1D, the function $f_{ij}(x)=
\Theta(R_i+R_j-|x|)$ is recovered as a convolution of the 
`molecular volume' $\Theta(R_i-|x|)$ with delta functions,
\begin{equation}
\fl f_{ij}(x)={1 \over 2}\int \rmd x'\,\big[\Theta(R_i-|x-x'|)
\delta(R_j-|x'|)+\Theta(R_j-|x-x'|)\delta(R_i-|x'|)\big],
\label{convo1D}
\end{equation}
which points to the use of normalized spherical delta function shells 
as `molecular surface' weight functions to complement $\eta({\bf r})$.

In 3D the minimal set of weight functions with range $R_i$ and $R_j$
to recover the Mayer function spherical step at $|{\bf r}|=R_i+R_j$, as in
\eref{convo1D} for the 1D case, requires a scalar and a vector weight
functions to define two sets of weighted densities:
\begin{equation}
n_i({\bf r})={1\over{4\pi R_i^2}}\int \rmd{\bf r}'\,\delta(|{\bf r}'|-R_i)
\rho_i({\bf r}+{\bf r}'),
\label{esc}
\end{equation}
and
\begin{equation}
{\bf v}_i({\bf r})={1\over{4\pi R_i^2}}\int \rmd{\bf r}'\,\delta(|{\bf
r}'|-R_i)\rho_i({\bf r}+{\bf r}'){{{\bf r}'}\over {R_i}},
\label{vec}
\end{equation} 
for each molecular species ($i=1,\dots,m$) present in the system with
density $\rho_i({\bf r})$. The structure of the density functional
$\Phi[\rho]$ at quadratic order in $\eta({\bf r})$, $n_i({\bf r})$
and ${\bf v}_i({\bf r})$ is uniquely determined to recover the 
leading terms in \eref{virial} and \eref{c2virial},
\begin{eqnarray}
\fl \Phi[\rho]=\int \rmd{\bf r}\,\bigg(\eta({\bf r})\sum_{i=1}^mn_i({\bf r})+
2\pi\sum_{i,j=1}^mR_iR_j(R_i+R_j) [n_i({\bf r})n_j({\bf r})-
{\bf v}_i({\bf r})\cdot{\bf v}_j({\bf r})]\bigg) \nonumber \\
+ \Or(\rho^3),
\label{virialFMT}
\end{eqnarray}   
 
The original proposal for the FMT \cite{Yasha} is consistent with
\eref{virialFMT} and, with the guideline of the scaled particle theory,
recovers the full Percus-Yevick (PY) approximation for the direct 
correlation function of a bulk HS fluid. According with an general 
rule emanating from the FMT structure, the free energy density 
functional has as many additive terms as the space dimension,
\begin{equation}
\Phi[\rho]=\sum_{i=1}^D \Phi_i^{(D)}[\rho].
\label{Phi}
\end{equation}
For 3D hard spheres the three terms proposed by Rosenfeld were
\begin{eqnarray}
\fl \Phi_1[\rho] = -\int \rmd{\bf r}\,\log[1-\eta({\bf r})]
\sum_{i=1}^mn_i({\bf r}),
\label{Phi1} \\
\fl \Phi_2[\rho] = 2\pi\sum_{i,j=1}^m R_iR_j(R_i+R_j)
\int \rmd{\bf r}\,{{n_i({\bf r})n_j({\bf r})-{\bf v}_i({\bf r})
\cdot{\bf v}_j({\bf r})}\over{1-\eta({\bf r})}},   
\label{Phi2} \\
\fl \Phi_{3,\rm o}[\rho] = 8\pi^2\sum_{i,j,k=1}^m R_i^2R_j^2R_k^2
\int \rmd{\bf r}_i\,n_i({\bf r}){{{1\over 3}n_j({\bf r})n_k({\bf r})-
{\bf v}_j({\bf r})\cdot{\bf v}_k({\bf r})}\over{[1-\eta({\bf r})]^2}}.
\label{Phi3y}
\end{eqnarray}
All the complexity arising from the mixture of
$m$ different HS species is reduced to the evaluation of
$\eta({\bf r})$ in \eref{eta} and three moments
$X^{(s)}({\bf r})=\sum_i R_i^s X_i({\bf r})$,
with $s=0,1,2$, of the averaged densities $X_i=n_i,{\bf v}_i$.  
Thus, the theory 
may be applied even to polydisperse systems with a continuous 
distribution of the molecular radius \cite{ignacio}.

For one dimensional (1D) hard rods the exact free energy DF, as
found by Percus \cite{Percus}, is recovered from just the first term
in \eref{Phi1}, with the obvious translation of $\eta({\bf r})$ and 
$n_i({\bf r})$ to 1D. For hard disks in two dimensions (2D), Rosenfeld 
\cite{Yasha2D} proposed two terms with structures similar to
\eref{Phi1}--\eref{Phi2}. The results obtained with this approximation
for a monocomponent HS system were very good (better than those
obtained with any previous DF approximation) for problems like 
determining the profiles of a HS fluid against a hard wall, thus 
showing that the non-local
dependence obtained with the geometric measures \eref{eta}, \eref{esc}
and \eref{vec} is a better representation of the sharp dependence of the
Mayer function in the irreducible kernels of \eref{virial}. 
However, the study of the HS crystal as a density distribution made
of narrow peaks at the positions of the crystal lattice, which
had been successfully achieved with the WDA \cite{ADA,WDA}
and other DF approximations \cite{Baus}, gave fully unphysical results
with a negatively diverging free energy in the limit of infinitely
narrow peaks. The problem was related to the overlap of three 
delta-function shells in \eref{Phi3y}, giving rise to integrable
divergences. These singularities are harmless in the evaluation of 
the bulk fluid free energy and direct correlation function, but they
notoriously show up when
the free energy is evaluated for delta-function density distributions.
In the case of a monocomponent HS system, the pathology of the original
FMT was analyzed in terms of the dimensional crossover of the DF from 
3D hard spheres to 1D hard rods, as well as the zero dimension (0D)
limit, i.e.\ narrow cavities which cannot contain more than one molecule
\cite{PRE,RC}. A new version of $\Phi_3[\rho]$ was proposed to
solve the problem \cite{PRL}, with the new ingredient of a 
2-rank tensor-weighted density $\mathcal{T}_i$, with cartesian components 
\begin{equation}
T_{i}^{(\alpha,\beta)}({\bf r})=
{1 \over{4 \pi R_i^2}} \int \rmd{\bf r}'\,\delta(|{\bf r}'|-R_i)
\rho_i({\bf r}+{\bf r}'){{r'_{\alpha} r'_{\beta}}\over {R_i^2}},  
\label{ten}
\end{equation} 
$\alpha, \beta=x,y,z$ and where the index $i$ is included to provide 
the obvious generalization to HS mixtures. This tensor-weighted 
density allows to write $\Phi_3[\rho]$ in such a way that it 
vanishes for any 1D distribution of HS, i.e.\ for any $\rho({\bf r})=
\delta(x) \delta(y) \rho_1(z)$, which should be fully equivalent
to a 1D system of hard rods with density distribution $\rho_1(z)$.
The combined forms of $\Phi_1[\rho]$ and $\Phi_2[\rho]$ give the exact
DF form in this limit, but the original term $\Phi_3[\rho]$ in \eref{Phi3y}
spoils the agreement. The new version of the FMT for the monocomponent 
HS system reproduces the exact 1D limit and gives an excellent description 
of the HS crystal, solving many of the qualitative problems of other DF
approximations \cite{PRL}. 

The extension of this new version of FMT to HS mixtures amounts to taking
\begin{equation}
\Phi_{3}[\rho]= 12\pi^2\sum_{i,j,k=1}^m R_i^2 R_j^2 R_k^2\int \rmd{\bf r} 
\,{{\varphi_{ijk}({\bf r})} \over{[1-\eta({\bf r})]^2}},
\label{Phi3n}
\end{equation}
with 
\begin{equation}
\varphi_{ijk}({\bf r})={\bf v}_i\cdot\mathcal{T}_j\cdot{\bf v}_k-
n_j{\bf v}_i\cdot{\bf v}_k-\Tr[\mathcal{T}_i\mathcal{T}_j\mathcal{T}_k]+
n_j\Tr[\mathcal{T}_i\mathcal{T}_k],
\label{f3}
\end{equation}
in terms of the tensorial contractions and trace of ${\bf v}_i({\bf r})$ 
and $\mathcal{T}_i({\bf r})$, for each component. 

However, such a direct extension 
to mixtures fails to reproduce the exact 1D limit. The reason for this 
failure, the modification in the DF structure required to achieve that
limit and the effect of such modifications in typical problems are to
be analyzed here. In the next section, the origin and effects of the 
spurious divergences in the FMT density functional are analyzed in terms 
of the triplet direct correlation function and a new FMT density
functional for HS mixtures is proposed to recover the exact 1D limit.
This new version, however, patches out this effect at the expense of
creating new weaker but potentially harmful singularities which preclude
using this functional for a free minimization.
In section 3 we compare the results of this new version with those
obtained with \eref{Phi3n} and \eref{f3}, and discuss the pros and
cons of using each. We conclude in section 4 that the functional
\eref{Phi3n}, \eref{f3}, despite its defects, provides the best balance
between accuracy and computational simplicity, and that it seems
impossible to go any further without modifying the basic
structure of FMT.

\section{FMT for HS mixtures}

The triplet direct correlation functions obtained for HS in three 
dimensions with any FMT density functional has a peculiar structure, 
very different from
that obtained with the WDA or similar approximations. The most relevant
form appears in the contribution at zero order in a density expansion,
from the functional derivative of $\Phi_{3}[\rho]$ in \eref{Phi},
which has the generic form
\begin{eqnarray}
\fl {{\delta^3 \Phi_3[\rho]}\over{\delta\rho_i({\bf r}_1)
\delta\rho_j({\bf r}_2)\delta\rho_k({\bf r}_3)}} =&
\int \rmd{\bf r}\,\delta(R_i-|{\bf r}_1'|)
\delta(R_j-|{\bf r}_2'|)\delta(R_k-|{\bf r}_3'|)
K_{ijk}({\bf r}_1',{\bf r}_2',{\bf r}_3') \nonumber \\
&+ \Or(\rho),
\label{c3FMT}
\end{eqnarray}
where $i,j,k=1,\dots,m$. The radii of the HS placed at
${\bf r}_1$, ${\bf r}_2$ and ${\bf r}_3$ are $R_i$, $R_j$ and $R_k$ 
respectively, and for each possible choice of the three indices,
$i,j,k$, there is a generic function $K_{ijk}$ of the relative 
vectors ${\bf r}_l'\equiv {\bf r}_l-{\bf r}$, $l=1,2,3$, whose 
moduli are restricted to be equal to the respective HS radii by the 
three delta-function shells. Higher order terms in the density expansion 
of $ c^{(3)}_{ijk}$, as well as the contributions to the triplet direct 
correlation function from the terms $\Phi_1[\rho]$ and $\Phi_2[\rho]$
in \eref{Phi1} and \eref{Phi2}, have one or more delta-function
shells substituted by spherical step functions (i.e.\ 
$\Theta(R_i-|{\bf r}_1'|)$, etc.) coming from the functional 
derivative of $\eta({\bf r})$. These contributions are always regular 
smooth functions, which cannot produce spurious divergences but
cannot reproduce either the exact step-like dependence of 
$c_{ijk}({\bf r}_1,{\bf r}_2,{\bf r}_3)=-f_{ij}(r_{12})
f_{jk}(r_{23}) f_{ki}(r_{31}) + \Or(\rho)$, in terms of the 
Mayer functions, $f_{ij}(r)=\Theta(R_i+R_j-r)$, and the 
relative distances $r_{12}=|{\bf r}_1-{\bf r}_2|$.

\begin{figure}
\begin{center}
\includegraphics*[height=2.2in, angle=-90]{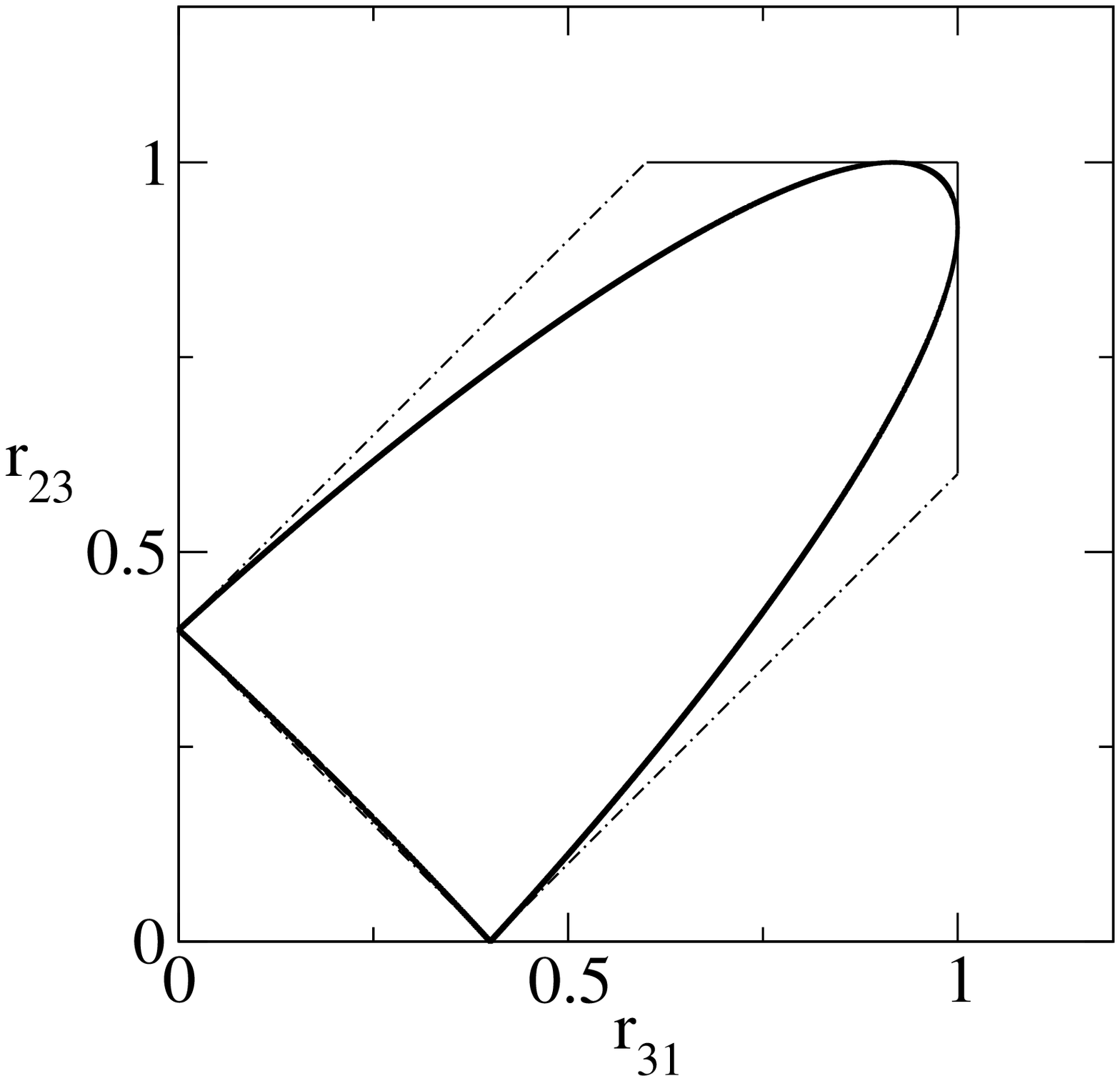} 
\includegraphics*[height=2.2in, angle=-90]{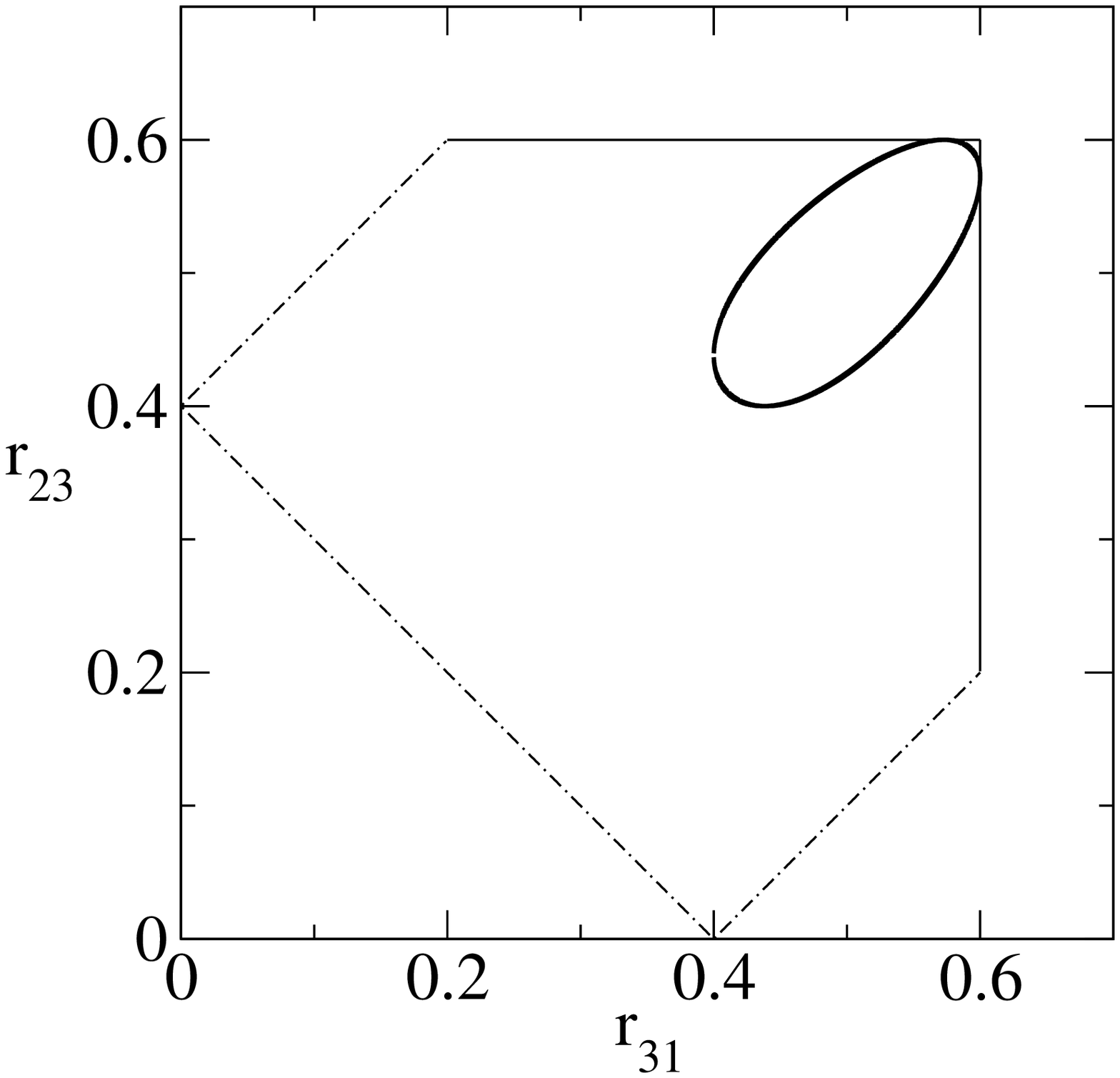}
\end{center}
\centerline{\hspace*{1cm} (a) \hspace*{5cm} (b)}
\caption{\label{fig1}Boundaries in the plane $(r_{23},r_{31})$,
for the given value of $r_{12}=0.4$. The dash-dotted lines 
give the triangle inequalities $|r_{23}-r_{31}| \leq r_{12} \leq
r_{23}+r_{31}$, for the possible values of these variables.
The solid thick curves give the regions of  
common overlap between three spheres placed at
${\bf r}_1$, ${\bf r}_2$ and ${\bf r}_3$ while the 
solid thin lines give the boundaries of
the simultaneous pairwise overlap. The HS radii are 
of equal size ($R_i=R_j=R_k=0.5$) (a) and of 
very different size ($R_i=R_j=0.5$ and $R_k=0.1$) (b).}
\end{figure}

\begin{figure}
\begin{center}
\vspace*{2mm}
\includegraphics*[height=2.8in, angle=-90]{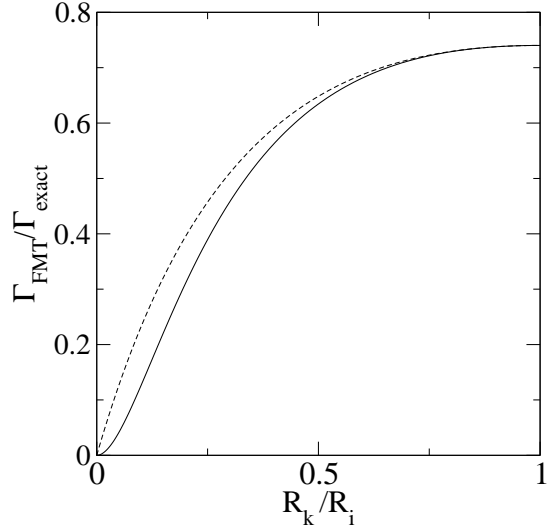}
\end{center}
\caption{\label{fig2}Ratio between $\Gamma^{(3)}_{\rm FMT}=
4\pi^4R_i^2R_j^2 R_k^2$ and $\Gamma^{(3)}_{\rm exact}$ 
[equation \eref{Gamma}] for $R_j=R_i\ge R_k$ (solid line)
and for $R_j=R_k\le R_i$ (dashed line).}
\end{figure}

It is only from the overlap of the three delta-function shells
in \eref{c3FMT} that the $c^{(3)}_{ijk}$ term in a FMT may be 
able to reproduce such a sharp step dependence. But the same overlap is 
also the origin of the spurious divergences which may invalidate 
the use of the FMT for density distributions with strong 
inhomogeneities. The crucial point to balance these two aspects is the 
choice of the kernel function $K_{ijk}({\bf r}_1',{\bf r}_2',
{\bf r}_3')$, which is precisely the difference between different 
versions of the FMT. The common aspect of all versions is that 
the range of the triplet correlation function, 
$c^{(3)}_{ijk}({\bf r}_1,{\bf r}_2,{\bf r}_3)$,
is restricted to points ${\bf r}_1$, ${\bf r}_2$ and ${\bf r}_3$
such that the three spheres of radii $R_i$, $R_j$ and $R_k$,
respectively centered at those points, have a common overlap.
This geometrical boundary is always inside the exact range
imposed by the Mayer functions product (which extends as far as
pairwise overlaps between the three spheres exist) but, 
particularly in the case of mixtures of HS with very different sizes, 
it is much more restrictive than the exact one. In \fref{fig1}
we represent the boundaries in the plane $(r_{23},r_{31})$
for a given value of $r_{12}$, both for HS of equal size (a) 
and of very different size (b). The total hyper-volume obtained
by integration over ${\bf r}_{2}$ and ${\bf r}_{3}$ in the 
region with non-zero values for \eref{c3FMT} may be obtained 
as a function of the HS radii as $\Gamma_{\rm FMT}^{(3)}=4\pi^4
R_i^2R_j^2 R_k^2$, while the equivalent integral for the exact
product of Mayer functions gives
\begin{eqnarray}
\fl \Gamma_{\rm exact}^{(3)}={{16\pi^2}\over{9}} \bigg[
R_i^3 R_j^3 + R_j^3 R_k^3 +R_k^3 R_i^3 \nonumber \\
+ 3 R_i^2 R_j^2 R_k^2 \left(3+\frac{R_i+R_j}{R_k}+ 
\frac{R_j+R_k}{R_i}+\frac{R_k+R_i}{R_j} \right) \bigg].
\label{Gamma}
\end{eqnarray}
The difference between $\Gamma_{\rm exact}^{(3)}$ and 
$\Gamma_{\rm FMT}^{(3)}$ corresponds to the FMT `lost-cases' 
described in the 0D approach to the FMT \cite{RC}, i.e.\ those 
configurations of three delta-function peaks which cannot give
contribution to $\Phi_3[\rho]$ in the FMT scheme, but 
which should give a contribution to the exact free energy
excess. When the three HS radii are equal the FMT contribution
covers about $74\%$ of the exact hyper-volume, so that
the role of the `lost-cases' is relatively mild, and its effect
on the overall accuracy of the DF approximation for usual
applications could be reduced by an appropriate choice of the 
kernel function in \eref{c3FMT}, which could increase the contribution
of the points inside the FMT boundary to overcome the lack of 
contribution of the missing configurations. However, the ratio 
between $\Gamma^{(3)}_{\rm FMT}$ and $\Gamma^{(3)}_{\rm exact}$ goes 
to zero in a mixture of very asymmetric HS, whenever the size
of one species goes to zero while keeping the size of the other two.
This is shown in \fref{fig2} and corresponds to the effect which was
obvious for the particular case in \fref{fig1}(b): the
contribution of $\Phi_3[\rho]$ associated to spheres of very 
different size shrinks to a very small part of the exact contribution 
to the triplet direct correlation function. If the kernel function
in \eref{c3FMT} is fixed to reproduce a given equation of state
for the bulk fluid mixture, the contribution 
of order $\rho_i \rho_j \rho_k$ coming from  $\Phi_3[\rho]$
would require an enormous artificial enhancement of the weight of 
those configuration within the FMT boundary, in order to compensate 
the lack of contribution from the missing configurations.
The use of such an approximation for inhomogeneous density distributions
could lead to quite an erroneous evaluation of the excess free 
energy, with spurious sensitivity to the density distribution 
of the small spheres. This could be of particular relevance in 
the DF study of depletion forces in colloidal particles 
for strongly inhomogeneous density distributions.  

Beside the problem of its reduced range, the mathematical 
form \eref{c3FMT} leads to the most serious problem of the 
FMT: the presence of spurious divergences at the boundary 
of non-zero values, arising from the tangency of two or 
the three spherical shells. This is clear if the integral over 
${\bf r}$ is transformed into integrals over the three moduli 
$r_1'$, $r_2'$ and $r_3'$, which can be directly performed 
thanks to the three delta functions in \eref{c3FMT}. The boundary 
of non-zero values is therefore given by the existence of a triple 
overlap between the three spheres; since at that boundary the 
vectors ${\bf r}_1'$, ${\bf r}_2'$ and ${\bf r}_3'$ are geometrically 
fixed, integration leads to
\begin{equation}
\fl
{{\delta^3 \Phi_3[\rho]}\over{\delta\rho_i({\bf r}_1)
\delta\rho_j({\bf r}_2)\delta\rho_k({\bf r}_3)}}=
{{4 R_i R_j R_k\,K_{ijk}({\bf r}_1',{\bf r}_2',{\bf r}_3')}\over
{[Z(r_{12},r_{23},r_{31},R_i,R_j,R_k)]^{1/2}}}
+ \Or(\rho).
\label{c3FMTz}
\end{equation}
The function $Z(r_{12},r_{23},r_{31},R_i,R_j,R_k)$
is the polynomial
\begin{eqnarray}
\fl Z = (R_i^2  r_{23}^2 + R_j^2 R_k^2)( r_{12}^2 - r_{23}^2+ 
r_{31}^2) + (R_j^2  r_{31}^2 + R_k^2 R_i^2)( r_{23}^2 - 
r_{31}^2+ r_{12}^2) \nonumber \\
\lo{\displaystyle +(R_k^2  r_{12}^2 +}R_i^2 R_j^2)( r_{31}^2 - 
r_{12}^2+ r_{23}^2) - R_i^4 r_{23}^2 - R_j^4 r_{31}^2 - 
R_k^4 r_{12}^2 - r_{12}^2 r_{23}^2 r_{31}^2
\end{eqnarray}
arising from the Jacobian of the transformation. This polynomial
vanishes over the whole boundary. The divergences of the terms in 
\eref{c3FMT} and \eref{c3FMTz} may only be avoided if the kernel 
also vanishes over the boundary. The version of the FMT based on 
the exact 0D limit for narrow cavities of arbitrary shape \cite{RC}, 
fulfills precisely such a condition with a function 
$K_{ijk}({\bf r}_1',{\bf r}_2',{\bf r}_3')$ proportional to
$[Z(r_{12},r_{23},r_{31},R_i,R_j,R_k)]^{1/2}$. This removes
the divergences in \eref{c3FMTz} while at the same time saves 
the step-like dependence between the inside and the outside of the 
FMT boundary. However, such a kernel function is not separable in powers 
of the cartesian components of the vectors ${\bf r}_l'$, so that 
the free energy cannot be evaluated in terms of simple weighted 
densities, like \eref{barra}, making it very costly to use it for 
practical applications.

The original FMT version \eref{Phi3y} corresponds to building
a kernel function on the basis of the bulk fluid
direct correlation (from the integration of $c^{(3)}_{ijk}$
with respect to one of the positions, as in \eref{c2virial})
and to get an easy 
computation of $\Phi[\rho]$ in terms of the scalar and vector weight 
densities \eref{esc} and \eref{vec}. In that case the kernel
function has to be decomposable in terms which are linear  on each
cartesian component of the vector variables. Within these
constraints the unique choice was \eref{Phi3y}, which is
equivalent to taking in \eref{c3FMT} and \eref{c3FMTz} the kernel
\begin{equation}
K_{ijk}={1 \over{24 \pi}} \left( 1 
- {{{\bf r}_1'\cdot {\bf r}_2'}\over {R_i \ R_j}}
- {{{\bf r}_2'\cdot {\bf r}_3'}\over {R_j \ R_k}}
- {{{\bf r}_3'\cdot {\bf r}_1'}\over {R_k \ R_i}} \right).
\label{Ky}
\end{equation}
However, such a choice for the kernel function
does not eliminate any of the divergences at the 
boundaries of the tangent spherical shells.  The strongest effect 
appears when $R_i=R_j=R_k$ and ${\bf r}_1={\bf r}_2={\bf r}_3$, so 
that the three spherical shells overlap in their whole surface. In 
that case \eref{Ky} takes the value $-(12 \pi)^{-1}$,
and there is a diverging negative contribution to the free energy.
The divergence is integrable and it contributes to give the exact
$\rho^3$ term in free energy \eref{virial} for uniform bulk fluid, 
but in a density distribution with narrow peaks (like in a 0D cavity
or in a crystal phase) the negative divergence goes to
$\Phi_3[\rho]$ in \eref{Phi3y} and the approximation 
becomes useless. Also, in the crossover from the 3D hard spheres
to the 1D system,
there is a spurious contribution of the divergence in $\Phi_3[\rho]$,
as given in \eref{Phi3y}, while the exact 1D density functional
would be recovered with a vanishing $\Phi_3[\rho]$ for such 
density distributions.

The FMT version \eref{Phi3n} and \eref{f3} was developed 
to recover the exact 1D limit for monocomponent HS systems 
\cite{PRL}. It corresponds to choosing the kernel
\begin{equation}
\fl K_{ijk} = {1 \over{16 \pi}} \left[ 
\left( 1 - {{{\bf r}_1'\cdot {\bf r}_2'}\over {R_i \ R_j}} \right)
\left( 1 - {{{\bf r}_2'\cdot {\bf r}_3'}\over {R_j \ R_k}} \right)
\left( 1 - {{{\bf r}_3'\cdot {\bf r}_1'}\over {R_k \ R_i}} \right) 
 - {{ [{\bf r}_1'\cdot ({\bf r}_2' \times {\bf r}_3')]^2}
\over{ (R_i \ R_j \ R_k)^2}} \right].
\label{Kn}
\end{equation}
The presence of a quadratic dependence in the cartesian 
components of the vector variables implies the need of
the tensor weighted density, with rank two, defined in
\eref{ten}, together with the scalar and vector weighted densities
\eref{esc} and \eref{vec} used in \eref{Phi3y}. The kernel has
two terms with different geometrical structure: one goes with
the triple vector product ${\bf r}_1'\cdot ({\bf r}_2' \times {\bf r}_3')$
which becomes zero whenever the three vectors ${\bf r}_1'$, ${\bf r}_2'$
and ${\bf r}_3'$ become coplanar. This happens always at the boundaries 
of the triple overlap of the three spherical shells in \eref{c3FMT}
so that this contribution to the kernel alone would get rid of
all the spurious divergences generated by the vanishing of the denominator
in \eref{c3FMTz}. Unfortunately, a kernel function based only on this
term would not reproduce the correct behavior for $c^{(2)}(r,\rho)$ in
a homogeneous fluid. The other contribution in \eref{Kn} becomes zero
whenever two of the three vectors ${\bf r}_l'$ become equal, which 
suppresses the strongest divergences of 
$c^{(3)}({\bf r}_1,{\bf r}_2,{\bf r}_3)$
produced when two delta functions shells, of equal radii, overlap their
full surfaces. In the monocomponent case, this is enough to make
the full contribution of $\Phi_3[\rho]$ to vanish in the 1D limit,
because three spherical shells of equal radii and with the centers
along a straight line, cannot have a triple overlap unless two of them
share the same center. 

Thus, in a monocomponent HS system the FMT version with \eref{Phi3n}
is free of the strongest divergences in 
$c^{(3)}({\bf r}_1,{\bf r}_2,{\bf r}_3)$,
produced when ${\bf r}_l={\bf r}_m$ and when the three points
${\bf r}_1,{\bf r}_2,{\bf r}_3$ are along a straight line 
(so that the spherical shells have a common axis).  However, 
there are still weak divergences at the boundary points
between overlapping and non-overlapping shells. These divergences
would appear in $\Phi[\rho]$ for very peculiar
density distributions, with three delta function peaks, normalized 
to contain less than one molecule among the three, and separated 
by distances such that the surfaces of molecules centered at
the three sites would intersect in a single point. Of course
such a density distribution would be very unlikely to appear 
unless the external potential acting on the system has been
tailored on purpose, and in that case the error made by this
FMT version would be of the opposite sign to the error made 
by \eref{Phi3y} in the 0D limit. Instead of a negative divergence 
of the free energy (from the negative sign of the kernel \eref{Ky}
in such a limit), there would be a positive divergence of the 
free energy because the kernel \eref{Kn} is always equal or larger
than zero for coplanar vectors. A positive divergence in $\Phi_3[\rho]$
for a very peculiar density distribution would  produce little harm
because  the minimization process of the free energy in the density 
functional formalism would just avoid such distribution functions.
The problem is much more innocuous than a negative divergence of the
free energy, which would invalidate the formal use of $\Phi[\rho]$
to minimize with respect to any possible function $\rho({\bf r})$.

\begin{figure}
\begin{center}
\vspace*{2mm}
\includegraphics*[width=3.0in, angle=0]{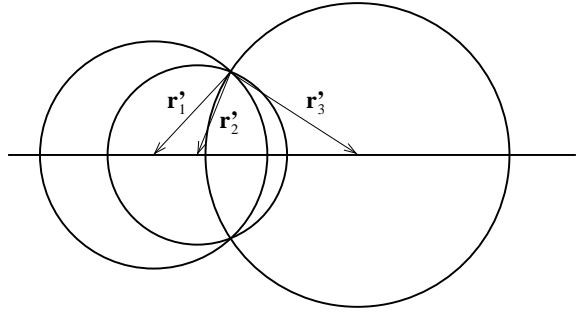}
\caption{\label{fig3}Three coaxial spheres of different 
radii overlapping
along a circumference. This situation is impossible for spheres of
the same radius, unless two of them overlap their whole surfaces.}
\end{center}
\end{figure}

In the case of HS mixtures with different radii, the kernel \eref{Kn}
is less efficient than in the monocomponent case in the regularization of
the FMT.  The exact 1D limit would still require that $\Phi_3[\rho]$
vanishes whenever all the points with non-zero density are along a 
straight line, but the kernel \eref{Kn} does not vanish in the
case of three coaxial spheres with a triple overlap along a circumference
(as illustrated in \fref{fig3}).
To recover the exact 1D density functional from a mixture of 3D
hard spheres confined along a straight line requires a kernel
$K_{ijk}({\bf r}_1',{\bf r}_2',{\bf r}_3')$ which vanishes whenever
the three vectors (with respective moduli $R_i$, $R_j$ and $R_k$)
go from a common origin to three points along a straight line.
The simplest geometrical construction with this requirement is
\begin{eqnarray}
\fl
K_{ijk}=
\frac{({\bf r}'_1 \times {\bf r}'_2+ {\bf r}'_2 \times {\bf r}'_3+
{\bf r}'_3 \times {\bf r}'_1)}{16 \pi R_i R_j R_k} 
\cdot\left[
\left(\frac{{\bf r}'_1\cdot{\bf r}'_2}{R_i+R_j}+
\frac{{\bf r}'_1\cdot{\bf r}'_3}{R_i+R_k}\right)
\frac{({\bf r}'_3 \times {\bf r}'_2)}{R_k \ R_j} \right. +
\nonumber \\
\fl
+
\left(\frac{{\bf r}'_2\cdot{\bf r}'_3}{R_j+R_k}+
\frac{{\bf r}'_2\cdot{\bf r}'_1}{R_j+R_i}\right)
\frac{({\bf r}'_1 \times {\bf r}'_3)}{R_i \ R_k} 
%\nonumber \\
+ \left.
\left(\frac{{\bf r}'_3\cdot{\bf r}'_1}{R_k+R_i}+
\frac{{\bf r}'_3\cdot{\bf r}'_2}{R_k+R_j}\right)
\frac{({\bf r}'_2 \times {\bf r}'_1)}{R_j \ R_i}
\right].
\label{newkernel}
\end{eqnarray}
The way to obtain this kernel is to multiply the vector
\begin{equation}
{\bf r}'_1 \times {\bf r}'_2+ {\bf r}'_2 \times {\bf r}'_3+
{\bf r}'_3 \times {\bf r}'_1=({\bf r}'_3-{\bf r}'_2)\times
({\bf r}'_1-{\bf r}'_2),
\end{equation}
which obviously vanish whenever the three vectors are coaxial,
by some linear combination of the three cross products in such
a way that the contribution to the triangle in the virial
expansion of the direct correlation function is recovered.
Such a thing cannot be accomplished if the coefficients are
just functions of the radii, but it can if we allow for a 
dependence on the dot products of the vectors.

The existence of cubic dependence on the vectors in  \eref{newkernel}
forces to use a new 3-rank tensor measure, namely 
$\mathcal{M}_i$, whose cartesian components are given by
\begin{equation}
M_{i}^{(\alpha,\beta,\gamma)}({\bf r})=
{1 \over{4 \pi R_i^2}} \int \rmd{\bf r}'\,\delta(|{\bf r}'|-R_i)
%\int {d{\bf r}' \over{4 \pi R_i^2}}\,\delta(|{\bf r}'|-R_i)
\rho_i({\bf r}+{\bf r}'){{r'_{\alpha} r'_{\beta} r'_{\gamma}}\over 
{R_i^3}}.
\label{rank3}
\end{equation}
In terms of this and the former measures, $\Phi_3[\rho]$ will
have the same shape as in \eref{Phi3n}, but with 
$\varphi_{ijk}({\bf r})$ in \eref{f3} replaced by 
$\varphi_{ijk}({\bf r})+\Delta\varphi_{ijk}({\bf r})$, where
\begin{equation}
\Delta\varphi_{ijk}({\bf r}) =
\frac{2R_j^2(R_i-R_k)}{R_i(R_i+R_j)(R_j+R_k)}
({\bf v}_i\cdot\mathcal{M}_j:\mathcal{T}_k
-{\bf v}_i\cdot\mathcal{T}_k\cdot{\bf v}_j),
\label{newf3}
\end{equation}
the symbol `:' denoting the contraction of two indices. An
important difference arises in this new version: the
coefficients multiplying the measures are no longer a product of 
powers of the radii, but more general rational functions of
them. This has the computational disadvantage of not allowing 
to express the functional in terms of a finite number of moments 
of the weighted densities, as in any previous FMT version.
This is particularly important in applications to polydisperse system,
where having moment dependent functionals dramatically simplifies
the calculations \cite{ignacio}.

A different and more formal drawback of the new kernel is the fact
that it cancels the most singular terms in the former kernel
(those that preclude it to recover the one-dimensional functional
for aligned configurations of the HS) at the expense of changing the sign
of the weak divergences for some non-aligned configurations, at
the boundary between overlapping and non-overlapping shells.
These divergences were always of positive sign for the kernel
\eref{Kn}, but for some configurations they become negative
with the kernel \eref{newkernel}.
This can be explicitly seen if we write
the kernel  \eref{newkernel} as 
\begin{equation}
K_{ijk}=\big[({\bf r}'_3-{\bf r}'_2)\times
({\bf r}'_1-{\bf r}'_2)\big]\cdot V_{ijk}({\bf r}'_1,{\bf r}'_2,
{\bf r}'_3),
\label{newk2}
\end{equation}
and now rotate ${\bf r}'_2$ slightly away of coaxiality while maintaining 
the three vectors in the same plane and with common origin (see \fref{fig3}).
If the displacement is given by the vector $\bepsilon$ ($|\bepsilon|
\ll R_j$) then $R_j=|{\bf r}'_2+\bepsilon|=R_j+2{\bf r}'_2\cdot
\bepsilon+\Or(|\bepsilon|^2)$, i.e.\ 
\begin{equation}
{\bf r}'_2\cdot\bepsilon=\Or(|\bepsilon|^2).
\label{eps1}
\end{equation} 
Replacing ${\bf r}'_2$ by ${\bf r}'_2+\bepsilon$ in \eref{newk2} yields
(remember that $K_{ijk}=0$ for coaxial vectors)
\begin{equation}
K_{ijk}=\big[({\bf r}'_3-{\bf r}'_1)\times\bepsilon\big]
\cdot V_{ijk}({\bf r}'_1,{\bf r}'_2,{\bf r}'_3)+\Or(|\bepsilon|^2).
\label{eps2}
\end{equation}
Clearly, if $\bepsilon$ fulfills \eref{eps1} and the r.h.s.\ of 
\eref{eps2} does not vanish (which happens, for instance, if
${\bf r}'_2$ and $\bepsilon$ are coplanar),
so does $-\bepsilon$, so the kernel may have both
signs, as stated.

A negative singularity is formally the worst defect 
a free energy density functional can have, because it implies that
the absolute minimum of the free-energy functional is minus
infinity, and it is reached for very singular density distributions.
In practice, however, this defect can be overcome with a restricted
minimization of the free energy within a functional family which 
does not include those singular distributions. Thus, the strong
negative divergence of the original FMT proposal \eref{Phi}-\eref{Phi3y}
for delta function density distributions, which precluded its use for 
crystals and narrow 0D cavities, does not interfere with its use
(and very good results) for systems with planar symmetry, when
the free energy is minimized with respect to planar density profiles
$\rho_i({\bf r})=\rho_i(z)$, independent of $x$ and $y$.
The remaining negative divergences in our new version 
\eref{eps2} are by far much weaker than those in \eref{Phi3y},
and they would be harmless for almost
any virtual application of density functional theory
with the variational minimization of the free energy
constrained to 
families of functions (uniform in one or two variables, like in
the adsorption at walls or in pores; periodic, like in freezing;
etc.). This notwithstanding, we will show in the next section that
the actual quantitative difference between this new functional
and the former one (equations \eref{Phi3n} and \eref{f3}) in
applications to standard problems is negligible, the former
functional being far simpler and easier to apply to polydisperse
mixtures.

\section{Results}

\begin{figure}
\begin{center}
%\vspace*{2mm}
\includegraphics*[width=3.0in, angle=0]{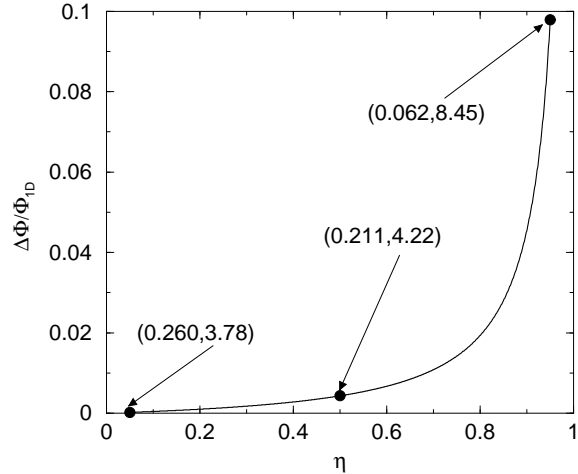}
\caption{\label{err-eta} Maximum, with respect to the radii ratio
$R_l/R_s$, and the relative packing fraction of the small segments, 
$x_s\equiv\eta_s/\eta$, of the relative error $\Delta\Phi/
\Phi_{\rm 1D}$ of the one-dimensional reduction of 
functional \eref{Phi3n} and \eref{f3} for a binary mixture, 
versus the total packing fraction, $\eta$, of the fluid.
The values of the pair $(x_s,R_l/R_s)$ are given for three 
points of the curve. It can be seen that the maximum error
occurs for $x_s\approx 0.2$ and $R_l/R_s\approx 4$ for most values
of $\eta$.}
\end{center}
\end{figure}

\begin{figure}
\begin{center}
%\vspace*{2mm}
\includegraphics*[width=2.5in, angle=0]{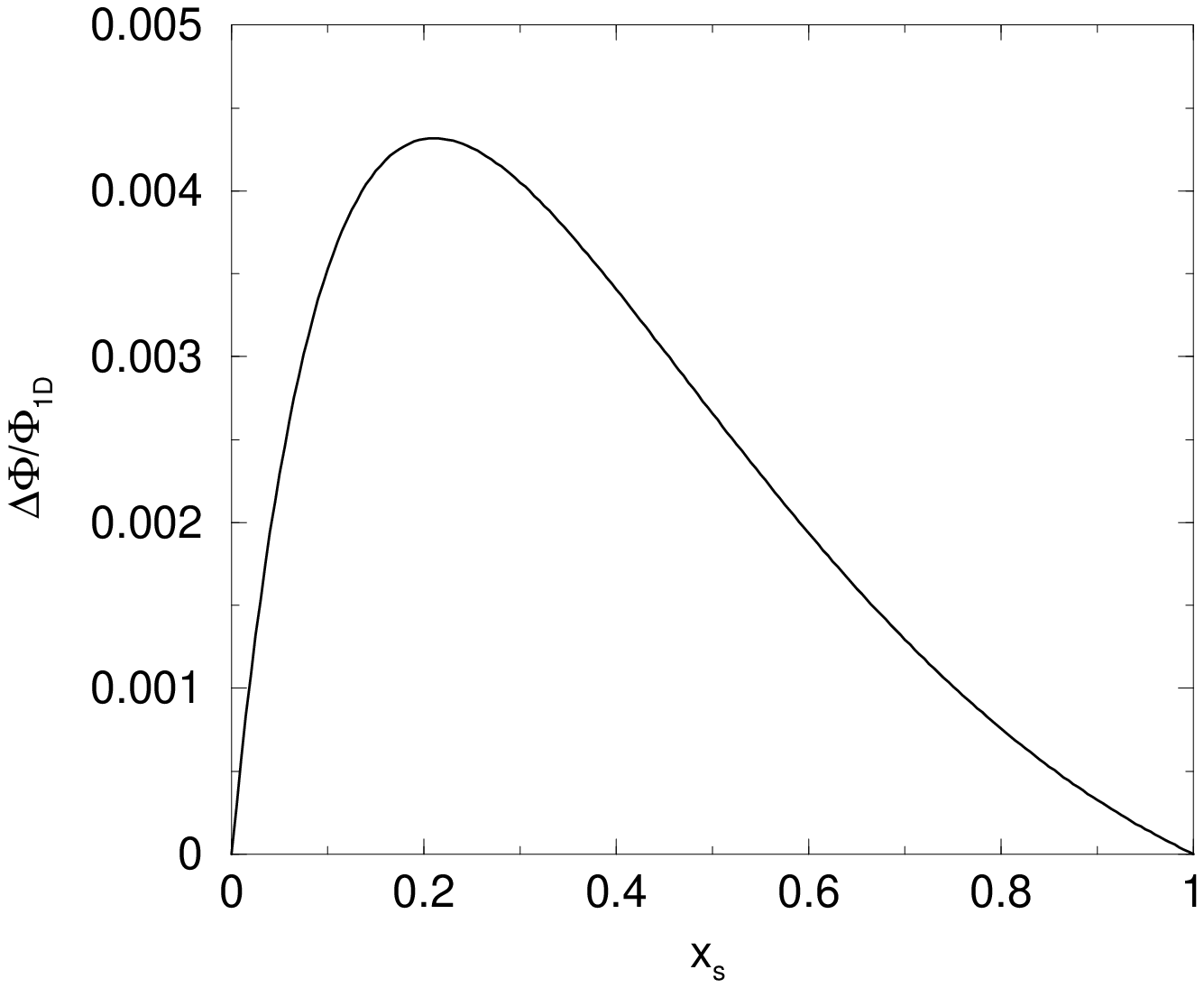}
\includegraphics*[width=2.5in, angle=0]{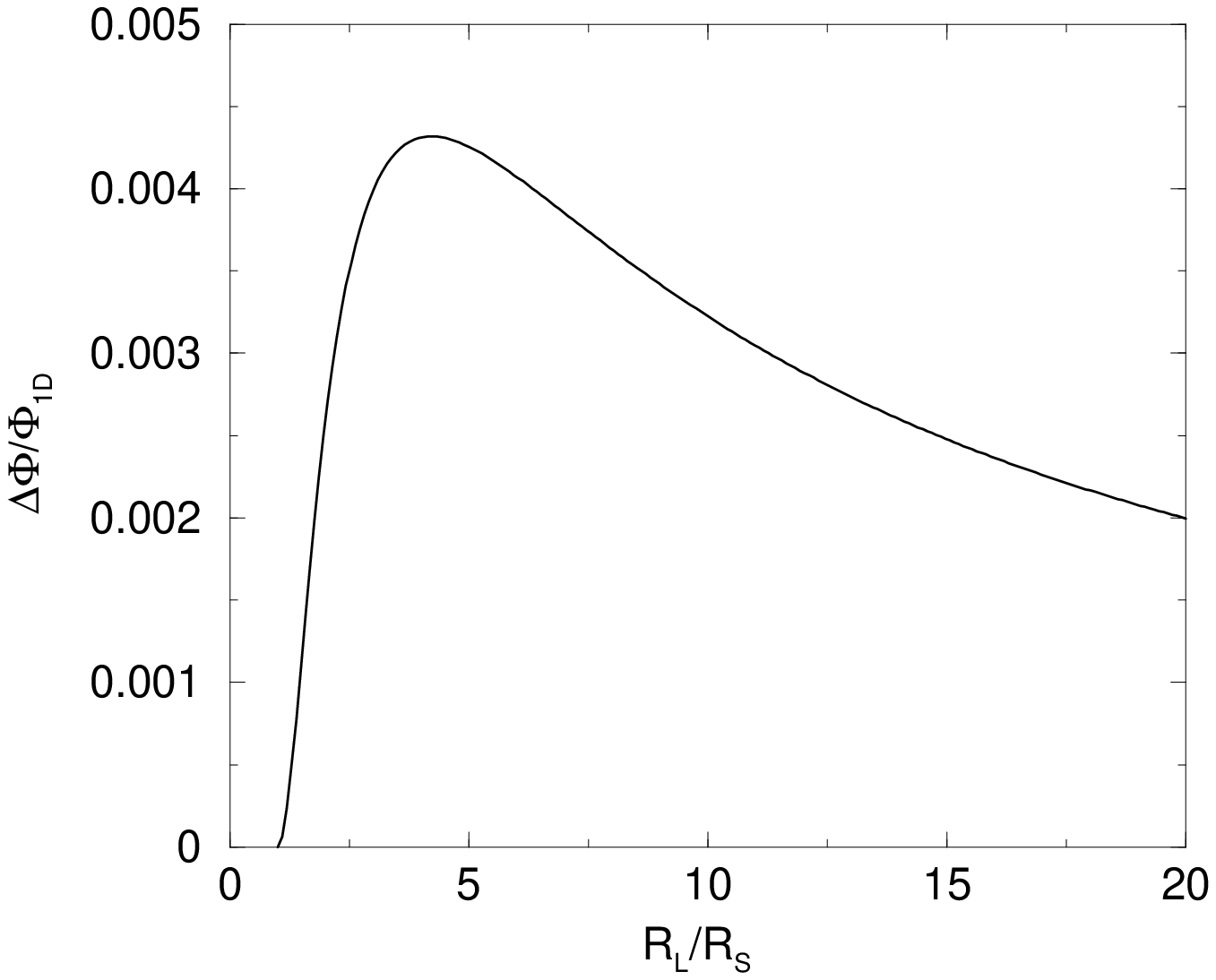}
\centerline{\hspace*{6mm} (a) \hspace*{57mm} (b)} 
\caption{\label{err-x} Relative error $\Delta\Phi/
\Phi_{\rm 1D}$ of the one-dimensional reduction of functional 
\eref{Phi3n} and \eref{f3} for a binary mixture at total
packing fraction $\eta=0.5$, versus the 
relative packing fraction of the small segments, $x_s\equiv
\eta_s/\eta$, for a radii ratio $R_l/R_s=4.22$ (a) as well as 
versus the radii ratio, $R_l/R_s$, for a value of $x_s=0.211$ 
(b).}
\end{center}
\end{figure}

As stated in the previous section, the functional \eref{Phi3n}
and \eref{f3} does not recover the exact free energy of a
uniform hard rod mixture when it is reduced to a one-dimensional 
density profile. The expression of the reduced functional is
rather cumbersome, but in order to estimate this deviation
we will simply evaluate the difference with respect to the exact
value for a binary mixture. Let us denote respectively $\Phi$ and 
$\Phi_{\rm 1D}$ the reduced free energy density (in units of 
$k_BT$) and the exact free energy density, and define
$\Delta\Phi=\Phi_{\rm 1D}-\Phi$. 
The relative error $\Delta\Phi/\Phi_{\rm 1D}$ has, for any
packing fraction $\eta$, an absolute maximum as a function of
the relative packing fraction of the small segments,
$x_s\equiv\eta_s/\eta$ and the radii ratio, $R_L/R_s$. This
maximum value is plotted vs.\ $\eta$ in \fref{err-eta}. In
\fref{err-x} we also plot the relative error 
$\Delta\Phi/\Phi_{\rm 1D}$ corresponding to $\eta=0.5$ both
vs.\ $x_s$ at fixed $R_L/R_s$ (a) and vs.\ $R_L/R_s$ at fixed
$x_s$ (b), for the values corresponding to the maximum 
relative error. Notice that this error remains smaller than 2\% up
to $\eta=0.8$, and it is never larger than 10\% even for 
packing fractions as high as $\eta=0.95$.

\begin{figure}
\begin{center}
%\vspace*{2mm}
\includegraphics*[width=2.5in, angle=0]{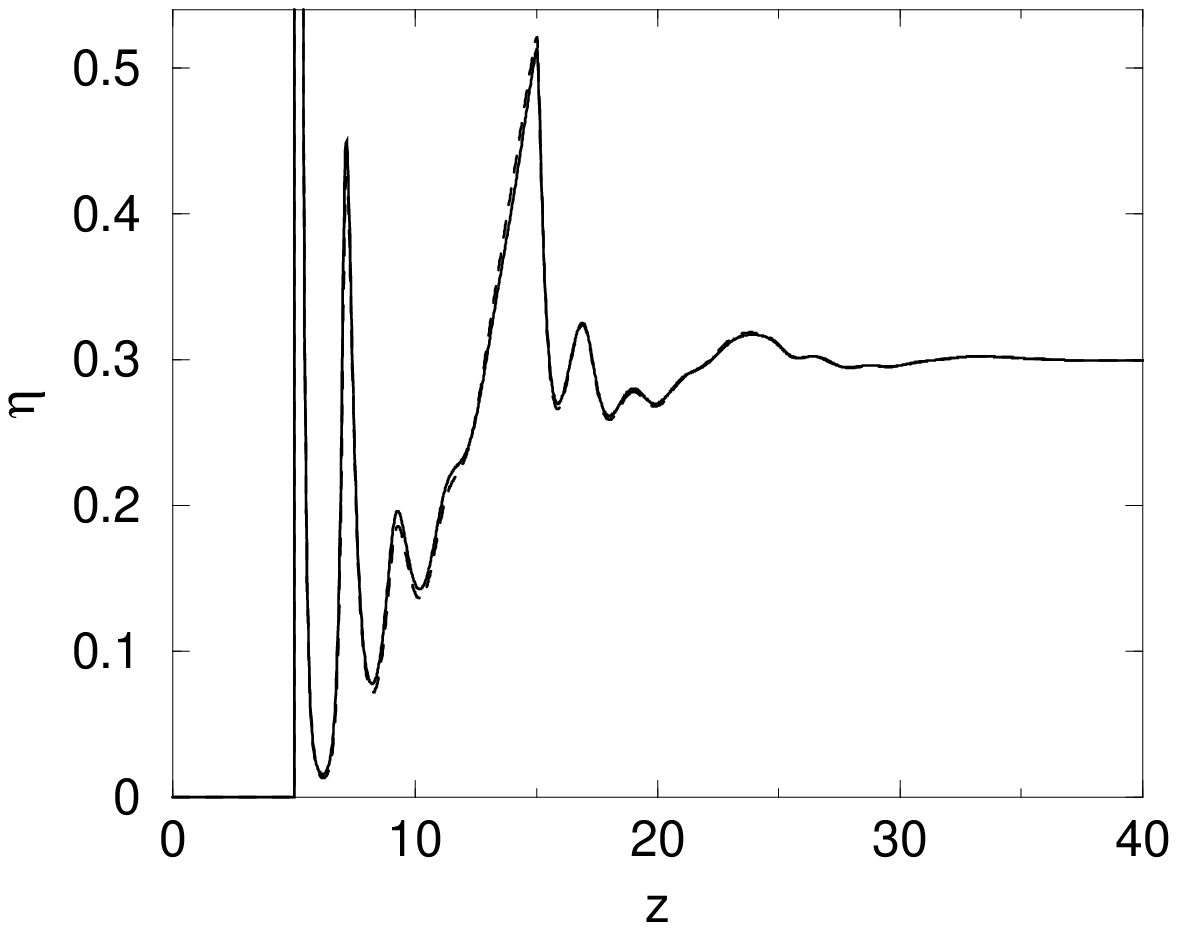}
\includegraphics*[width=2.5in, angle=0]{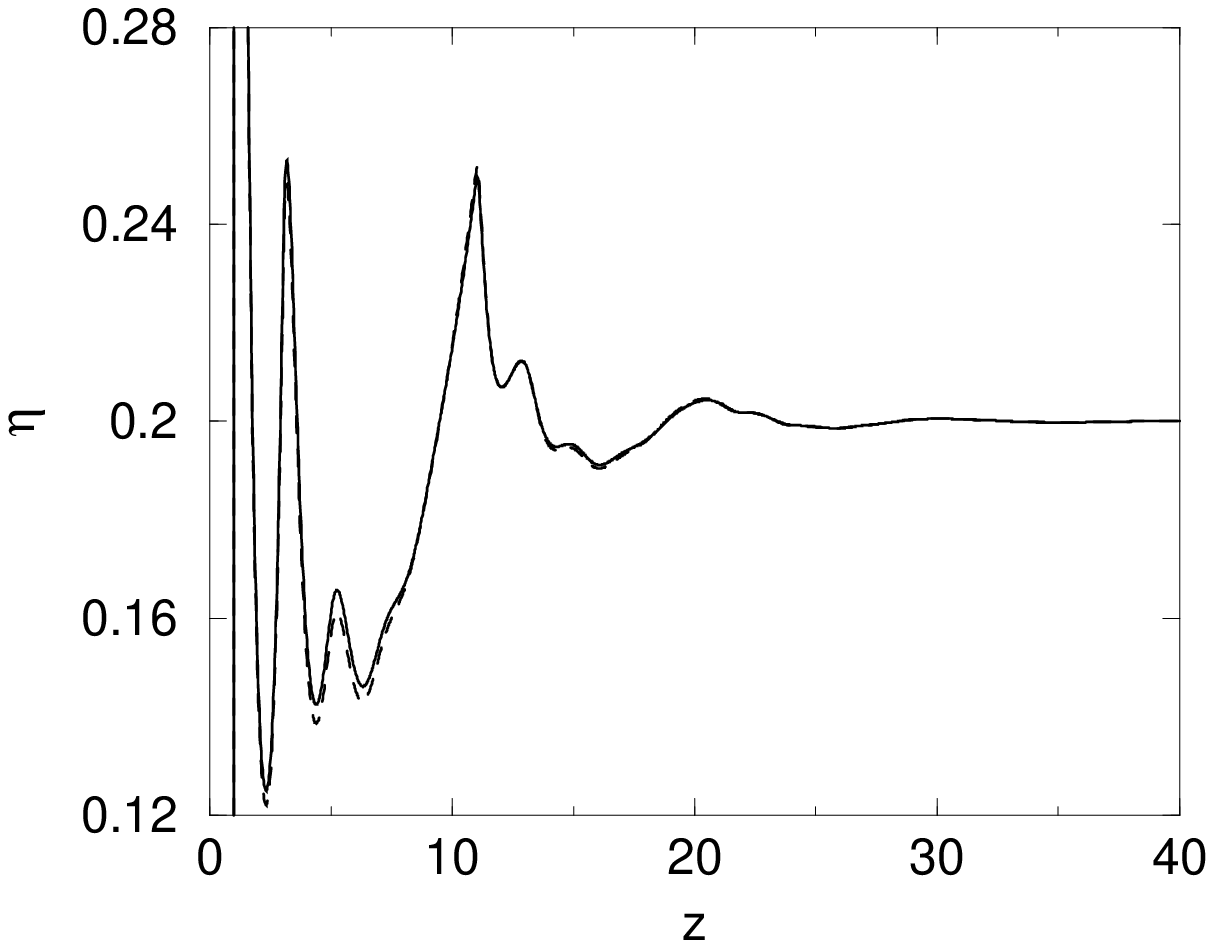}
\centerline{\hspace*{6mm} (a) \hspace*{57mm} (b)} 
\caption{\label{wall} Density profiles of a binary mixture of
hard spheres of radii $R_s=1$ and $R_l=5$ near a hard wall,
as obtained both with functional \eref{Phi3n}, \eref{f3} only
(dashed line) and with the correction \eref{newf3} (solid line).
The densities are multiplied by the volume of each type of
particle. The bulk packing fractions of the small and large 
spheres are,
respectively, $\eta_s=0.2$ and $\eta_l=0.3$. The difference
between the results of both functionals can hardly be seen in
the plotting scale.}
\end{center}
\end{figure}

\begin{figure}
\begin{center}
%\vspace*{2mm}
\includegraphics*[width=3.0in, angle=0]{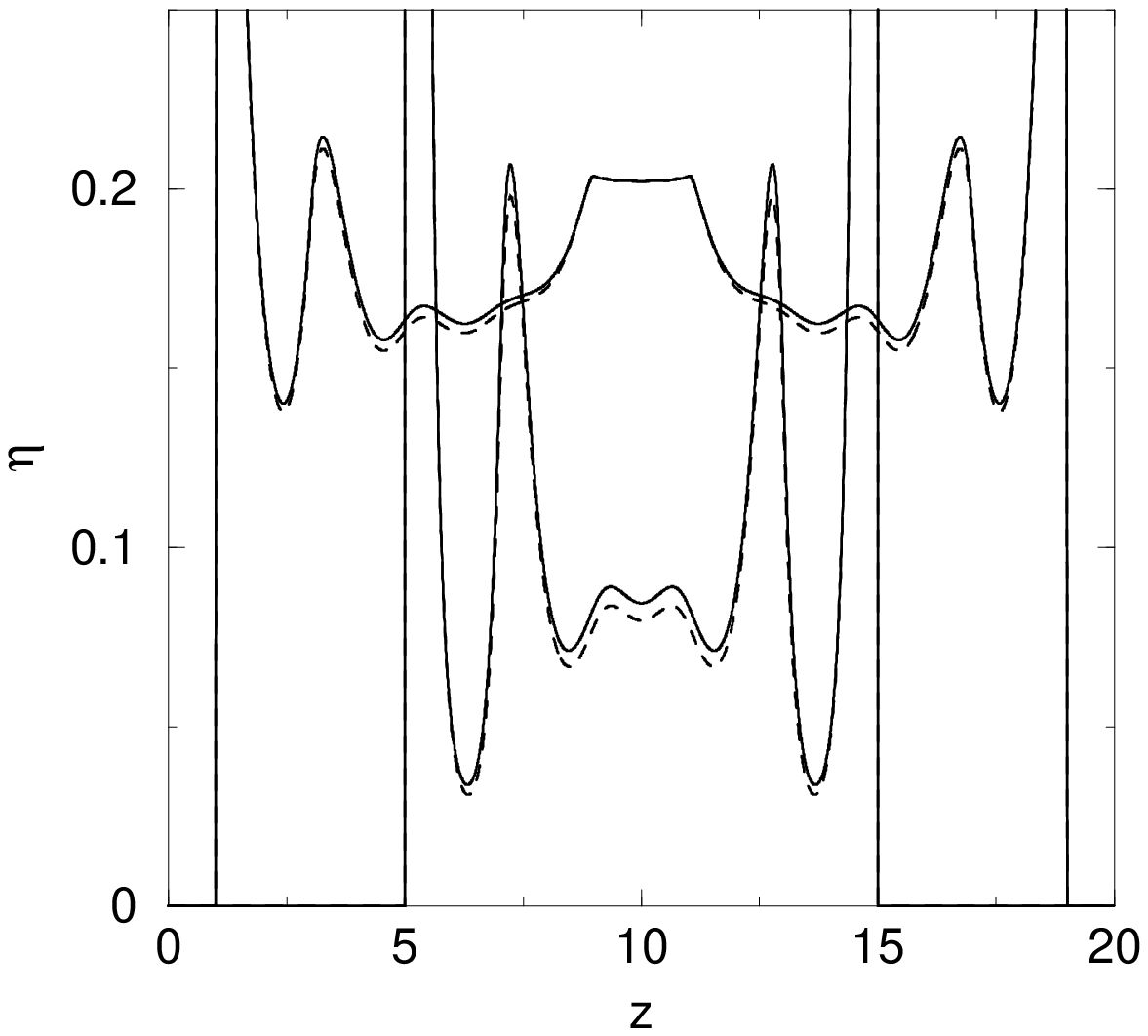}
\caption{\label{slit} The same as in \fref{wall} but for
a slit of length $20R_s$. The system is in chemical equilibrium
with a reservoir with packing fractions 
$\eta_s=0.2$ and $\eta_l=0.2$. Although the difference 
is now more visible, it is still negligible.}
\end{center}
\end{figure}

The conclusion we extract from these figures is that, despite
having an incorrect dimensional crossover to 1D, the functional
\eref{Phi3n}, \eref{f3} produces results of high accuracy in
this limit. A second test can be obtained by applying this
functional and the one with the correction \eref{newf3} to
obtain the density profiles of a binary mixture of hard spheres
both, near a hard wall and within a slit. The parameters of the
system have been chosen in order to obtain the most prominent 
differences. The results are shown in \fref{wall}, for 
the case of the hard wall, and \fref{slit} for the case of the 
slit. We see again that differences are negligible.

In view of these numerical tests we conclude that functional
\eref{Phi3n}, \eref{f3} is preferable to the corrected one, despite
its inability to recover the one-dimensional free energy exactly,
simply because its structure is far simpler and more suitable
to be applied to polydisperse mixtures (see for instance
\cite{ignacio}).

\section{Conclusions}

The main conclusion of the analysis we have performed on the
construction of a fundamental measure functional for mixtures of
hard spheres is that, as stated in the title, we appear to be `close to
the edge' of fundamental measure theory, in the sense that its 
internal structure seems  almost exhausted. There are two
main drawbacks that seem unavoidable within the present scheme,
both related to the structure of the third term $\Phi_3[\rho]$ in 
\eref{Phi}. The first problem is common to all the FMT versions and 
it concerns the existence of `lost cases', i.e.\
the fact that configurations of three spheres with pairwise overlap
but no triple overlap do not make any contribution to the
functional. This reflects in that the lowest order of $c_3$
is incorrectly predicted (it vanishes for the lost cases,
where it should still be 1). The obvious way to repair this
defect is to introduce two point measures joining two
`halves' of Mayer functions. This would be a qualitative 
departure from the FMT structure, which would recover the exact
$c_3$ at low density, but the increase in
computational complexity would be considerable. Another way
to circumvent this problem, while keeping the FMT structure and
low computational cost, would be to replace the delta
functions in the definition of $\Phi_3[\rho]$ by some other
functions having a `tail' extending beyond the radii of the
particles (and presumably vanishing when particles no longer
overlap). Although we cannot definitely exclude this possibility,
we have tried several functional forms without success. This
seems to be a too drastic change within the fundamental
measure scheme, so much that adding such tails would `break down'
the functional at some other point. In particular, the dimensional 
crossover (to the exact 1D limit and to 0D cavities, strongly
related to the density distribution in crystals) is a very stringent, 
hence fragile, requirement for an approximate functional, so virtually 
any modification can spoil it.

The existence of these lost cases forces to overweight the
other configurations in order to compensate for the nonexistent
ones and guarantee a reasonable equation of state. But as 
\fref{fig1} and \fref{fig2} show, this compensation
introduces a dramatic bias in the case of
mixtures of very dissimilar spheres. This may, at least in
principle, have important consequences for the depletion effect 
in these systems and thus may affect phase behaviour, at least
in its use to describe crystalline phases. Although in the application
to planar density profiles, the effect would be strongly reduced
by the averaging over the transverse directions.
 
The problem of the lost cases seems to be inherent to the
fundamental measure structure with a notable exception: parallel
hard parallelepipeds \cite{cubos}. The peculiar shape of these bodies makes
pairwise overlap and triple overlap to be equivalent conditions,
so there are no lost cases for this particular form of the
particles, and hence the lowest order in $c_3$ is exactly
recovered.

The second problem, also associated to $\Phi_3[\rho]$, depends
on the choice of the kernel function 
$K_{ijk}({\bf r}_1',{\bf r}_2',{\bf r}_3')$ in \eref{c3FMT},
which marks the difference between the FMT versions proposed
in the literature. The balance here comes between the computational
cost and the existence of spurious divergences (of positive or
negative sign) in the free energy. The original form
proposed by Rosenfeld factorizes in terms of the scalar
and vector weighted densities, but the negative divergences show 
up in the reduction to 1D and to 0D distributions. Simpler versions
(for the monocomponent case) using no kernel function and
only the scalar weighted density 
\cite{White} give much stronger divergences (showing up even at the 
bulk fluid direct correlation function $c(r,\rho_0)$) 
but always with positive sign, so that in the minimization of 
the free energy those configurations contributing to the 
divergences are avoided, e.g. creating spurious kinks in the 
density profiles of a wall-fluid interface, but without the
qualitative breakdown of \eref{Ky} for strongly 
confined systems.

  In contrast, the FMT version based on the exact reduction to the 
0D limit \cite{RC} would have no divergences of the free energy at all, 
but it has a non-separable kernel with enormous computational cost
(implying the multiple integrals over the densities at 
three different points). The proposal made in \cite{RC} and used by 
Groh and Mulder \cite{Groh} of taking 
$K_{ijk} \sim [{\bf r}_1'\cdot ({\bf r}_2' \times {\bf r}_3')]^2$
would also kill all the divergences
at the boundary between overlapping and non-overlapping spheres,
and be separable in terms of the scalar, vector and a rank-two
tensor weighted densities. However, this form gives a too smooth
boundary behaviour for \eref{c3FMT}, since the kernel
vanishes as the function $Z$ in the square root of denominator in 
\eref{c3FMTz}, while the step-like dependence would 
require a non-separable kernel proportional to
$K_{ijk} \sim |{\bf r}_1'\cdot ({\bf r}_2' \times {\bf r}_3')|$.
The alternative kernel \eref{Kn} and DF \eref{Phi3n}, 
proposed and successfully
used \cite{PRL} for monocomponent HS systems, achieved
the separability (in term of the same set of weighted densities)
and the sharpness (with the exact contribution to the triangle 
diagram for the bulk fluid), at the price of unleashing some
weak divergences of positive sign of the free energy 
for some peculiar configurations of non-aligned molecules.
The effect of those divergences would be extremely difficult
to observe and the quality of that DF approximation seems to
be limited mainly by the quality of the Percus-Yevick equation
of state for the bulk HS fluid, and the associated condition
of having a direct correlation function with the range of the 
hard core diameter \cite{Cancun}, which is also intrinsic to 
the FMT.

 The extension to HS mixtures imposes more severe restrictions,
since the form \eref{Kn} for HS of unequal size has 
(positive) divergences for peculiar aligned configurations
which were not present for equal sized HS and which give
stronger contributions than those of non-aligned configurations.
Their main drawback is to spoil the exact reduction
to the 1D limit, which was one of the most remarkable 
achievements for the monocomponent case.  We have 
introduced here a new kernel form \eref{newkernel}
which would recover that dimensional reduction for 
HS mixtures, at the price of introducing a new
rank-three tensorial weighted density and creating
some weak negative divergences for some non-aligned 
configurations. Besides, the radii-dependent
coefficients of this new kernel are not simply products of
powers of the
radii, but more complicated rational functions, so it turns out
to be computationally more involved in applications to, for
instance, polydisperse mixtures.  Other possibilities remain
for a candidate kernel, like squaring the one we propose,
conveniently adjusting the coefficients; however this forces the
introduction of very high rank tensor weights and removes the
discontinuity of $c_3$, thus creating both computational
and structural problems.         

   Altogether, the systematic improvement of the free energy
density functional within the FMT seems to be frustrated by
incompatible requirements, deeply rooted in the
structure of the theory. Thus, 
one has to choose a functional form on the basis of
its performance for the required practical use. The 
original proposal by Rosenfeld is quite accurate
for systems with planar symmetry, for which the 
improvement obtained with the newer versions is purely marginal.
The FMT version \eref{Phi3n}, with a moderate increase
of the computational cost due to the presence of the tensorial
weighted density \eref{ten}, would give similar results
as  \eref{Phi3y} for planar systems but it is also good 
for crystalline phases or other strongly confined density
distributions. In the 1D limit it does not recover the 
exact DF for arbitrary distributions, but making a rather small
error unless the 1D system is very close to the close
packing limit. The extra computational cost of the FMT version
\eref{newf3}, with a rank-three tensor and non-additivity of the 
weighted densities in terms of a few moments of the radii,
would only be worth for problems in which the exact 1D 
reduction is crucial and the spurious negative divergences
of this free energy DF are made innocuous by the parametrization of
the density distribution. It seems very unlikely that these
two constrains appear in a practical problem, so that probably
\eref{Phi3n} is the best practical choice.

To conclude, we have to point that the problems and limitations
of the FMT density functionals, which we have 
explored here, should be considered in the right perspective,
as we are asking the FMT functionals to perform well in 
problems which would be out of question for any other
type of DF approximation. The exact (or nearly exact) reduction
from three dimensional density distributions to the 1D limit,
and the accurate behaviour in very narrow cavities, which
opens the description of the crystalline phases with unconstrained
density distributions, are fully out of range for the WDA or
other DF approximations. The presence of spurious divergences
in the FMT is the price to be paid for the sharpness of
its non-local dependence, compared with the blurred dependence
of the other DF approaches. The taming of those divergences
to make them innocuous has been the `holy grail' of the 
workers in the field, starting from Rosenfeld's choice
for the combinations of $n_i({\bf r})$ and ${\bf v}_i({\bf r})$
in \eref{Phi2} and \eref{Phi3y}. The `earthly' requirements of 
practical computability make that `grail' unreachable, but
the new versions explored here seem to be close enough
to it for most practical purposes, providing by far the best
density functional approximations for systems of HS mixtures.

\section*{Acknowledgments}

This work is part of the research projects PB97-1223-C02-01
(DGESIC), BFM2001-1679-C03-02 (DGI) and BFM2000-0004 (DGI) of
the Ministerio de Ciencia y Tecnolog\'{\i}a (Spain). Yuri
Mart\'{\i}nez-Rat\'on acknowledges financial support from a 
postdoctoral grant of the Direcci\'on General de Investigaci\'on 
of the Consejer\'{\i}a de Educaci\'on de la Comunidad de 
Madrid (Spain).


\section*{References}

\begin{thebibliography}{99}

\bibitem{Yasha} Rosenfeld Y 1989 \PRL {\bf 63} 980 \\
	\dash 1993 \JCP {\bf 98} 8126 \\
	\dash 1994 \PR E {\bf 50} R3318 \\
	\dash 1996 \JPCM {\bf 8} 9287 \\
	\dash 1996 \JPCM {\bf 8} L795

\bibitem{others}  Kierlik E and Rosinberg M L 1990 \PR A {\bf 42} 3382 \\
	\dash 1991 \PR A {\bf 44} 5025 \\
	Phan S, Kierlik E, Rosinberg M L, Bildstein B and Kahl G 1993
	\PR E {\bf 48} 618

\bibitem{White} Gonz\'alez A, White J A and Evans R 1997 \JPCM {\bf 9} 2375
 
\bibitem{cubos} Cuesta J A 1996 \PRL {\bf 76} 3742 \\
	Cuesta J A and Mart\'{\i}nez-Rat\'on Y 1997 \PRL {\bf 78} 3681 \\
	\dash 1997 \JCP {\bf 107} 6379 \\
	Mart\'{\i}nez-Rat\'on Y and Cuesta J A 1998 \PR E {\bf 58} R4080 \\
	\dash 1999 \JCP {\bf 1111} 317

\bibitem{schmidt}  Schmidt M 1999 \PR E {\bf 60} R6291 \\
	\dash 2000 \PRL {\bf 85} 1934 \\
	\dash 2000 \PR E {\bf 63} 010101(R) \\
	\dash 2001 \PR E {\bf 63} 050201(R)

\bibitem{Hansen} Hansen J P and McDonald I R 1986 {\em Theory of
	Simple Liquids} 2nd ed.\ (London: Academic Press)

\bibitem{ADA} Tarazona P 1985 \PR A {\bf 31}, 2672

\bibitem{WDA} Curtin W A and Ashcroft N W 1989 \PR A {\bf 32} 2909

\bibitem{PY}  Percus J K and Yevick G J 1958 \PR {\bf 110} 1 \\
	Thiele E 1963 \JCP {\bf 39} 474 \\
	Wertheim M S 1963 \PRL {\bf 10} 321

\bibitem{Percus} Percus J K 1976 {\em J.\ Stat.\ Phys.} {\bf 15} 505
 
\bibitem{ignacio} Pagonabarraga I, Cates M E and Ackland G J 2000 
	\PRL {\bf 84} 911

\bibitem{Yasha2D} Rosenfeld Y 1990 \PR A {\bf 42}, 5978

\bibitem{Baus} Lutsko J F and Baus M 1990 \PR A {\bf 41}, 6647 \\
	Tejero C F, Ripoll M S and P\'erez A 1995 \PR E {\bf 52} 3632

\bibitem{PRE} Rosenfeld Y, Schmidt M, L\"owen H and Tarazona P 1997
	\PR E {\bf 55} 4245

\bibitem{RC} Tarazona P and Rosenfeld Y 1997 \PR E {\bf 55} R4873 \\
	\dash 1999 {\em New Approaches to Problems in Liquid State 
	Theory} ed C Caccamo et al.\ (Dordrecht: Kluwer) p 293

\bibitem{PRL} Tarazona P 2000 \PRL {\bf 84} 694
 
\bibitem{Groh} Groh B and Mulder B 2000 \PR E {\bf 61} 3811 \\
	Groh B 2000 \PR E {\bf 61} 5218

\bibitem{Cancun} Tarazona P 2002 {\em Physica} A {\bf 306} 243

\end{thebibliography}
\end{document}